\documentclass[aps,twocolumn,floatfix,superscriptaddress,preprintnumbers,showpacs,showkeys]{revtex4}
\usepackage{amssymb}
\usepackage{amsmath}
\usepackage{amsfonts}
\usepackage{latexsym,graphicx,epsfig}
\usepackage{color}

\newcommand{\bea}{\begin{eqnarray}}
\newcommand{\ena}{\end{eqnarray}}
\newcommand{\nn}{\nonumber\\}
\newcommand{\be}{\begin{equation}}
\newcommand{\en}{\end{equation}}

\newcommand{\ed}{\end{document}}

\newcommand{\la}{\langle}
\newcommand{\ra}{\rangle}

\newcommand{\slp}{p\kern-5pt/}

\begin{document}
                
\hfill MITP/15-109 (Mainz) 

\title{
Semileptonic decays $\Lambda_c^+ \to \Lambda \ell^+ \nu_\ell\,\,(\ell=e,\mu)$
in the covariant quark model and comparison with the new absolute 
branching fraction measurements of Belle and BESIII}

\author{Thomas Gutsche}
\affiliation{
Institut f\"ur Theoretische Physik, Universit\"at T\"ubingen,
Kepler Center for Astro and Particle Physics, 
Auf der Morgenstelle 14, D-72076, T\"ubingen, Germany}

\author{Mikhail A. Ivanov}
\affiliation{Bogoliubov Laboratory of Theoretical Physics, 
Joint Institute for Nuclear Research, 141980 Dubna, Russia}

\author{J\"{u}rgen G. K\"{o}rner}
\affiliation{PRISMA Cluster of Excellence, Institut f\"{u}r Physik, 
Johannes Gutenberg-Universit\"{a}t, 
D-55099 Mainz, Germany}

\author{Valery E. Lyubovitskij}
\affiliation{
Institut f\"ur Theoretische Physik, Universit\"at T\"ubingen,
Kepler Center for Astro and Particle Physics, 
Auf der Morgenstelle 14, D-72076, T\"ubingen, Germany}
\affiliation{ 
Department of Physics, Tomsk State University,  
634050 Tomsk, Russia} 
\affiliation{Mathematical Physics Department, 
Tomsk Polytechnic University, 
Lenin Avenue 30, 634050 Tomsk, Russia} 

\author{Pietro Santorelli}
\affiliation{
Dipartimento di Fisica, Universit\`a di Napoli
Federico II, Complesso Universitario di Monte Sant' Angelo,
Via Cintia, Edificio 6, 80126 Napoli, Italy} 
\affiliation{
Istituto Nazionale di Fisica Nucleare, Sezione di Napoli, 
80126 Napoli, Italy} 

\today

\begin{abstract}

We present precise theoretical predictions for the
absolute branching fractions of $\Lambda_c^+ \to \Lambda \ell^+ \nu_\ell\,(\ell=e,\mu)$ 
decays in the covariant confined quark model. 
This study is motivated by two recent and accurate measurements of 
the absolute branching fractions of $\Lambda_c^+ \to p K^- \pi^+$ 
and $\Lambda_c^+ \to \Lambda e^+ \nu_e$ by the Belle Collaboration 
at the KEKB and by the BESIII Collaboration at the BEPCII. 
Our predictions for the branching fractions are consistent with 
both experimental results. We also provide detailed numerical results 
for differential decay distributions and polarization observables. 

\end{abstract}

\pacs{12.39.Ki,13.30.Eg,14.20.Jn,14.20.Lq}
\keywords{relativistic quark model, light and heavy baryons,
decay rates and asymmetries}

\maketitle

\section{Introduction}

In 2013 the Belle Collaboration at KEKB~\cite{Zupanc:2013iki} 
reported on the first model-independent measurement of the branching fraction
${\rm Br}(\Lambda_c^+ \to p K^- \pi^+) = (6.84 \pm 0.24^{+0.21}_{-0.27})\%$.
This measurement significantly improved the precision of the absolute branching
fractions of other $\Lambda_c^+$ decay modes and of $b$-flavored hadrons
involving the $\Lambda_c^+$ state. In particular, using the Belle result 
the Particle Data Group~\cite{Agashe:2014kda} updated their average 
for the branching fractions of the exclusive semileptonic modes of 
the $\Lambda_c^+$ to
\bea
{\rm Br}(\Lambda_c^+ \to \Lambda e^+ \nu_e) &=& (2.9 \pm 0.5) \% \,, 
\nonumber\\
{\rm Br}(\Lambda_c^+ \to \Lambda \mu^+ \nu_\mu) &=& (2.7 \pm 0.6) \% 
\,. 
\ena  
A few months ago the BESIII Collaboration reported on the first absolute 
measurement of the branching ratio of 
$\Lambda_c^+ \to \Lambda e^+ \nu_e = (3.63 \pm 0.38 ({\rm stat}) 
\pm 0.20 ({\rm syst})) \%$~\cite{Ablikim:2015prg}. 
The current upper limit, given in the 2015 Review of Particle Physics,
agrees with the lower limit of the BESIII result.  
The new data calls for a detailed theoretical analysis of the 
$\Lambda_c^+ \to \Lambda \ell^+ \nu_\ell$ ($\ell = e, \mu$) process.  

In this paper we present a comprehensive study of 
the semileptonic decay $\Lambda_c^+ \to \Lambda \ell^+ \nu_\ell$ 
in the covariant confined quark 
model~\cite{Ivanov:1996fj}-\cite{Gutsche:2015mxa}. 
We calculate the total rate, 
differential decay distributions and polarization effects. 
The paper is organized as follows. In Sec.~II, we briefly review
the helicity formalism for the $\Lambda_{c} \to \Lambda$ transition 
which was discussed in detail e.g. in Ref.~\cite{Gutsche:2015mxa}. 
In Sec.~III we discuss the dynamics of the current--induced 
$\Lambda_{c} \to \Lambda$ transitions in terms of the covariant
confined quark model and present our numerical results.
Finally, in Sec.~IV we summarize and conclude our results. 

\section{Helicity formalism for the $\Lambda_{c} \to \Lambda$ transition}

First we define the matrix element $M_\mu^{V/A}(\lambda_{1},\lambda_{2}) 
= \langle B_2,\lambda_{2}|J_\mu^{V/A}|B_1,\lambda_{1}\rangle$ 
of the process 
$\Lambda_c^+(p_1)\to \Lambda^0(p_2) 
+ W^+_{\rm off-shell}(q)$, 
which can be expanded in terms of a complete set of invariants
\bea
M_\mu^V(\lambda_{1},\lambda_{2}) 
&=&  
\bar u_2(p_2,\lambda_{2})\bigg[F_1^V(q^2)\gamma_\mu-\frac{F_2^V(q^2)}{M_1}
  i\sigma_{\mu q}\nonumber\\
&+&\frac{F_3^V(q^2)}{M_1}q_\mu\bigg]u_1(p_1,\lambda_{1})\,,
\nonumber\\
\label{eq: invariants}\\
M_\mu^A(\lambda_{1},\lambda_{2}) &=& 
  \bar u_2(p_2,\lambda_{2})\bigg[F_1^A(q^2)\gamma_\mu-\frac{F_2^A(q^2)}{M_1}
  i\sigma_{\mu q}\nonumber\\ 
&+&\frac{F_3^A(q^2)}{M_1}q_\mu\bigg]\gamma_{5}u_1(p_1,\lambda_{1})\,,  
\nonumber
\ena
where $\sigma_{\mu q}=\frac{i}{2}(\gamma_\mu\!\!\not\!q\,  
- \not\! q \, \gamma_\mu)$ 
and  $q = p_1 - p_2$. The labels $\lambda_{i}=\pm \frac{1}{2}$ denote the 
helicities of the two baryons. In the present application $B_1=\Lambda_c$ and 
$B_2=\Lambda$.

Helicity amplitudes for the 
$\Lambda_c^+ \to \Lambda^0 + W^{+}_{\rm off-shell}$ transition 
are defined through the matrix elements as 
\be
H^{V/A}_{\lambda_2,\lambda_W} = 
M_\mu^{V/A}(\lambda_1,\lambda_2)\epsilon^{\dagger\,\mu}(\lambda_W)\,, 
\label{eq:hel_def}
\en
where there are four helicities for the $W^{+}_{\rm off-shell}$,
namely $\lambda_W=\pm 1,0\,(J=1)$ and $\lambda_W=0\,(J=0)$. See 
Fig.~\ref{fig:angles} for the definition of the kinematical variables 
for the decay process (the polar angles $\theta$, $\theta_B$ and the azimuthal 
angle $\chi$). The label $J=1,0$ 
denotes the two angular momenta of $W^{+}_{\rm off-shell}$ in its rest frame. 
Following the convention of Refs.~\cite{Korner:1989ve,Korner:1989qb} 
we distinguish the two $\lambda_W=0$ states and adopt
the notation $\lambda_W=0$ for $J=1$ and $\lambda_W=t$ for $J=0$ 
($t$ for temporal). From angular momentum conservation one has
$\lambda_{1}=\lambda_{2}-\lambda_{W}$. Using Eq.~(\ref{eq:hel_def}) 
one can derive explicit relations between the helicity amplitudes 
and the relativistic form factors~\cite{Gutsche:2013pp}-\cite{Gutsche:2015mxa} 
\bea
H_{+\frac12 t}^{V/A}&=&\frac{\sqrt{Q_\pm}}{\sqrt{q^2}}
  \bigg( M_\mp F_1^{V/A}\pm \frac{q^2}{M_1} F_3^{V/A}\bigg),
\nn
H_{+\frac12 +1}^{V/A}&=&\sqrt{2Q_\mp}
  \bigg(F_1^{V/A}\pm \frac{M_\pm}{M_1}F_2^{V/A}\bigg),
\nn
H_{+\frac12 0}^{V/A}&=&\frac{\sqrt{Q_\mp}}{\sqrt{q^2}}
  \bigg(M_\pm F_1^{V/A}\pm \frac{q^2}{M_1} F_2^{V/A}\bigg)\,, 
\label{eq:hel_inv}
\ena
where $M_\pm = M_1\pm M_2$ and $Q_\pm = M_\pm^2  - q^2$.
 
The total left--chiral helicity amplitude is defined by the composition
\bea
H_{\lambda_2,\lambda_W} = H_{\lambda_2,\lambda_W}^V 
- H_{\lambda_2,\lambda_W}^A \,, 
\ena 
where from parity or from an explicit calculation one has
\bea
H_{-\lambda_2,-\lambda_W}^V  = H_{\lambda_2,\lambda_W}^V,
\qquad
H_{-\lambda_2,-\lambda_W}^A = -H_{\lambda_2,\lambda_W}^A.
\ena 

The polarization observables to be discussed further on can be expressed
in terms of helicity structure functions given in terms of  bilinear 
combinations of helicity amplitudes. The definition of the structure
functions are collected in Table~\ref{tab:bilinears} 
(for more details see  Ref.~\cite{Gutsche:2015mxa}). 
\begin{table}[ht] 
\begin{center}
\caption{Definition of helicity structure functions} 
\def\arraystretch{1.5}
\begin{tabular}{l}
\hline
parity-conserving (p.c.) \\ 
\hline
${\cal H}_U   = |H_{+\frac12 +1}|^2 + |H_{-\frac12 -1}|^2$  \\
${\cal H}_L    = |H_{+\frac12\, 0}|^2 + |H_{-\frac12\, 0}|^2$  \\
${\cal H}_S    = |H_{+\frac12\, t}|^2 + |H_{-\frac12\, t}|^2$   \\
${\cal H}_{LT}  =   {\rm Re}\left(  H_{+\frac12 +1}  H_{-\frac12\, 0}^\dagger 
                         + H_{+\frac12\, 0} H_{-\frac12 -1}^\dagger \right)$ \\
 ${\cal H}_{ST} =  {\rm Re}\left( H_{+\frac12 +1} H_{-\frac12\,t}^\dagger 
                        + H_{+\frac12\,t} H_{-\frac12 -1}^\dagger \right)$ \\
${\cal H}_{SL} =  {\rm Re}\left(  H_{+\frac12\,0} H_{+\frac12\,t}^\dagger 
                        + H_{-\frac12\,0} H_{-\frac12\,t}^\dagger \right) $
\\[2ex]
\hline 
parity-violating (p.v.)  \\
\hline
${\cal H}_P   = |H_{+\frac12 +1}|^2 - |H_{-\frac12 -1}|^2$   \\

${\cal H}_{SL_P} =  {\rm Re}\left(  H_{+\frac12\,0} H_{+\frac12\,t}^\dagger 
                         - H_{-\frac12\,0} H_{-\frac12\,t}^\dagger \right)$ \\
${\cal H}_{L_P} = |H_{+\frac12\, 0}|^2 - |H_{-\frac12\, 0}|^2$  \\
${\cal H}_{S_P} = |H_{+\frac12\, t}|^2 - |H_{-\frac12\, t}|^2$   \\
${\cal H}_{LT_P} =   {\rm Re}\left(  H_{+\frac12 +1}  H_{-\frac12\, 0}^\dagger 
                           - H_{+\frac12\, 0} H_{-\frac12 -1}^\dagger \right)$ \\
${\cal H}_{ST_P} =  {\rm Re}\left(  H_{+\frac12 +1} H_{-\frac12\,t}^\dagger 
                          - H_{+\frac12\,t} H_{-\frac12 -1}^\dagger \right)$  
\\[2ex]
\hline
\end{tabular}
\label{tab:bilinears}
\end{center}
\end{table}

The helicity structure functions have definite parity properties as 
indicated in Table \ref{tab:bilinears}. The upper and lower halves of 
Table~\ref{tab:bilinears} list the parity-conserving (p.c.)
and parity-violating (p.v.) bilinear combinations of helicity amplitudes,
respectively, i.e. the p.c. and p.v. helicity structure functions are
linked to the products of
$VV$ and $AA$, and $VA$ and $AV$ currents, respectively.

The helicity amplitudes $H_{\lambda_2,\lambda_W}$ are a superposition of
vector and axial vector pieces and thus 
do not have definite parity properties. One can project back to the
vector and axial vector helicity amplitudes by defining the transversity
amplitudes~\cite{Kruger:2005ep} (see the relations 
in Ref.~\cite{Gutsche:2015mxa}).  

We do not list the full three-fold angular decay distribution of 
the cascade decay
$\Lambda_c \to \Lambda (\to p\pi^-) + \ell^+\nu_\ell$ 
but distill various asymmetries and polarization observables 
from the angular decay distribution as has been done
in~\cite{Gutsche:2015mxa}. 
When transcribing the results of~\cite{Gutsche:2015mxa} to the case
$\Lambda_c \to \Lambda \ell^+\nu_\ell$ treated here one has to take 
into account the necessary changes going from the $\ell^- \bar\nu_\ell$ 
to the $\ell^+ \nu_\ell$ final state lepton
configuration as discussed in~\cite{Gutsche:2015mxa}.

Next we list the expressions for the physical observables (width, 
differential rate, polarizations) in terms of the helicity structure 
functions: 

i) The normalized differential rate reads
\bea
\frac{d\Gamma}{dq^2} = 
\Gamma_{0} \frac{(q^2-m_\ell^2)^2 |\mathbf{p_2}|}{M_1^7q^2} \; 
{\cal H}_{\rm tot}
\label{eq:1-fold}
\ena
where 
\be
{\cal H}_{\rm tot} = {\cal H}_U\, +\, {\cal H}_L\, 
+\,  \delta_\ell \, \Big[
 {\cal H}_U\, +\, {\cal H}_L\,+\, 3\,{\cal H}_S
\Big] 
\en 
and where we define a helicity flip factor 
$\delta_{\ell}=m^{2}_{\ell}/(2q^2)$. 
In~(\ref{eq:1-fold}) we have introduced the Born term rate
\be
\Gamma_0=\frac{G^2|V_{cs}|^2 M_1^5}{192 \pi^3}\,.
\en
The rate $\Gamma_{0}$ represents the SM rate of the decay of a massive parent
fermion into three massless fermions, i.e. $M_1\neq 0$ and 
$M_2,m_\ell,m_{\nu_\ell}=0$, where $F^{V/A}_{1}=1$ and $F^{V/A}_{2,3}=0$.
The $q^{2}$--dependent factor multiplying $\Gamma_{0}$ in 
Eq.~(\ref{eq:1-fold}) is chosen such that the integral results in 1
for these mass and form factor settings. $V_{cs} = 0.986$ is the 
Cabibbo--Kobayashi--Maskawa matrix element. 

It is convenient to define partial rates $d\Gamma_{X}/dq^2$ and
$d\widetilde\Gamma_{X}/dq^2$ for the helicity nonflip~(nf) and
helicity flip~(hf) helicity structure functions ${\cal H}_{X}$ defined in
Table~\ref{tab:bilinears}. One has
\bea
\frac{d\Gamma_{X}}{dq^2}({\rm nf}) &=&
\Gamma_0 \,
\frac{(q^2-m_\ell^2)^2 |\mathbf{p_2}|}{M_1^7q^2} \, {\cal H}_X\,,
\nonumber\\
\frac{d\widetilde\Gamma_{X}}{dq^2}({\rm hf}) &=& \delta_{\ell}
\,\Gamma_0 \,
\frac{(q^2-m_\ell^2)^2 |\mathbf{p_2}|}{M_1^7q^2} \, {\cal H}_X\,. 
\ena
The partial rates can then be split into a helicity ``nf'' and helicity
``hf'' part according to
\be
\frac{d\Gamma_{X}}{dq^2}=\frac{d\Gamma_{X}}{dq^2}({\rm nf})
+\frac{d\widetilde\Gamma_{X}}{dq^2}({\rm hf}) \;.
\en
ii) The lepton-side forward-backward 
asymmetry defined by
\be
A_{FB}^{\ell}(q^2) 
= \frac{3}{4} \frac{{\cal H}_P\,-4\,\delta_\ell\,{\cal H}_{SL}}
{{\cal H}_{\rm tot}} 
\,. \label{eq:FB}
\en
iii) The convexity parameter $C_F(q^2)$ according to 
\be
C_F(q^2) = \frac{3}{4}\,(1\,-\,2\delta_\ell)\, 
   \frac{{\cal H}_U\,-\,2\,{\cal H}_{L}}{{\cal H}_{\rm tot}} .
\label{eq:CF}
\en
iv) Longitudinal $P^h_z(q^2)$ and transverse $P^h_x(q^2)$ 
polarizations of the daughter baryon $\Lambda$ 
\bea
P^h_z(q^2)  
&=& \frac{ {\cal H}_{P} + {\cal H}_{L_P} + \delta_{\ell}\,
({\cal H}_{P} + {\cal H}_{L_P} + 3{\cal H}_{S_P})}{{\cal H}_{\rm tot}}\,, 
 \nn[1.5ex]
P^h_x(q^2) &=& \frac{3\pi}{4\sqrt{2}}\frac{{\cal H}_{LT} 
+ 2\,\delta_{\ell}{\cal H}_{ST_P}}{{\cal H}_{\rm tot}} .
\label{eq:Pol_had}
\ena
v) Longitudinal $P^\ell_z(q^2)$ and transverse $P^\ell_x(q^2)$ 
polarizations of the charged lepton $\ell^+$ 
\bea
P^\ell_z(q^2)  &=& 
\frac{{\cal H}_U +{\cal H}_L-\delta_{\ell}\,( {\cal H}_U + {\cal H}_L 
+3{\cal H}_{S} )}{{\cal H}_{\rm tot}}\,, 
\nonumber \\[1.5ex]
P^\ell_x(q^2)  &=& -\frac{3\pi}{4\sqrt{2}}\sqrt{\delta_{\ell}}\,\,
\frac{{\cal H}_{P}+2\,{\cal H}_{SL}}{{\cal H}_{\rm tot}}
\label{eq:Pol_lept}
\ena
 When calculating the $q^{2}$--averages of the components of
$\vec P^{h}$and $\vec P^{\ell}$ one has to reinstate the common 
$q^{2}$--dependent factor
$(q^2-m_\ell^2)^2 |\mathbf{p_2}|/q^2$ in the numerator and denominator of
the right hand sides of Eqs.~(\ref{eq:Pol_had}) and (\ref{eq:Pol_lept}).

vi) The normalized azimuthal distribution is described by
\bea 
\widetilde W(\chi,q^2)=\frac{1}{2\pi}(1+\alpha_B \gamma \cos\chi) 
\ena 
where 
\be
\gamma(q^2)= \frac{3\pi^{2}}{16\sqrt{2}}
\frac{{\cal H}_{LT}+2\delta_{\ell}{\cal H}_{ST_{P}}}{{\cal H}_{\rm tot}}
=\frac{\pi}{4}P^{h}_{x}(q^{2})
\label{azipar1}
\en
and $\alpha_B$ is the asymmetry parameter in the decay $\Lambda \to p \pi^-$.

Next we define averages of the above observables: 
the mean forward-backward asymmetry $\la A_{FB} \ra$, 
the mean convexity parameter $\la C_F \ra$ and the hadronic 
$\la P^h_{x,z} \ra$ and 
leptonic  $\la P^\ell_{x,z} \ra $ polarization components, which are  
obtained from the nonflip and flip rates:  
\bea
\la A_{FB}^{\ell} \ra &=&\, \frac{3}{4}\, 
\frac{\Gamma_{P}-4\, \widetilde \Gamma_{SL}}{\Gamma_{\rm tot}} \,,\nonumber\\
\la C_F \ra &=&\,\frac{3}{4}\, 
\frac{\Gamma_{U}-2\Gamma_{L}-2\tilde\Gamma_{U}+4\tilde\Gamma_{L}}
{\Gamma_{\rm tot}} \,,\nonumber\\
\la P_{z}^h \ra &=&
\frac{\Gamma_{P} + \Gamma_{L_{P}} + \widetilde \Gamma_{P} 
+ \widetilde \Gamma_{L_{P}} + 
3\, \widetilde \Gamma_{S_{P}}}{\Gamma_{\rm tot}} \,,\nonumber\\
\la P_{x}^h \ra &=& \frac{3\pi}{4\sqrt{2}}\, 
\frac{\Gamma_{LT} + 2 \widetilde \Gamma_{ST_{P}}}{\Gamma_{\rm tot}} \,,
\nonumber\\
\la P_{z}^\ell \ra &=&
\frac{\Gamma_{U}+\Gamma_{L}-\widetilde\Gamma_{U}
-\widetilde\Gamma_{L}-3\, \widetilde\Gamma_{S}}{\Gamma_{\rm tot}} \,,
\nonumber\\
\la P_{x}^\ell \ra &=& - \frac{3\pi}{4\sqrt{2}}\,\sqrt{\delta_\ell} 
\frac{\Gamma_{P}+2\, \Gamma_{SL}}{\Gamma_{\rm tot}} \,,
\nonumber\\
\la \gamma \ra &=& \frac{\pi}{4} \, \la P_{z}^h \ra  \,. 
\ena

\section{The $\Lambda_c \to \Lambda_s$ transition
form factors in the covariant confined quark model}

We shall use the covariant confined quark model previously developed by us
to describe the dynamics of the current--induced 
$\Lambda_c = (c[ud])$ to
$\Lambda_s = (s[ud])$ transition 
(see Refs.~\cite{Gutsche:2012ze}-\cite{Gutsche:2015mxa}).  
The starting point of the model is an interaction Lagrangian
which describes the coupling of the $\Lambda_q$-baryon $(q=c,s)$ to
the relevant interpolating three-quark current. One has
\bea
{\cal L}^{\,\Lambda_q}_{\rm int}(x) 
&=&g_{\Lambda_q} \,\bar\Lambda_q(x)\cdot J_{\Lambda_q}(x) 
 + {\rm H.c.} \,, \label{eq:Lagr}\\
J_{\Lambda_q}(x) &=& \int\!\! dx_1 \!\! \int\!\! dx_2 \!\! \int\!\! dx_3 \, 
F_{\Lambda_q}(x;x_1,x_2,x_3) \nonumber\\
&\times& \epsilon^{a_1a_2a_3} \, q^{a_1}(x_1)\, u^{a_2}(x_2) 
\,C \, \gamma^5 \, d^{a_3}(x_3)\,.\nonumber  
\ena
The vertex function $F_{\Lambda_q}$ is chosen to be of the form
\be
F_{\Lambda_q}  \, = \, 
\delta^{(4)}(x - \sum\limits_{i=1}^3 w_i x_i) \;  
\Phi_{\Lambda_q}\biggl(\sum_{i<j}( x_i - x_j )^2 \biggr) 
\label{eq:vertex}
\en 
where $\Phi_{\Lambda_q}$ is a correlation function involving the three 
constituent quarks with coordinates $x_1$, $x_2$, $x_3$ and with 
masses $m_1$, $m_2$, $m_3$. The variable $w_i$ is defined by 
$w_i=m_i/(m_1+m_2+m_3)$ such that $\sum_{i=1}^3 w_i=1.$ 
The form factors describing the $\Lambda_c\to\Lambda_s$ transition
via the local weak quark current are calculated in terms of a two-loop 
Feynman diagram.
Due to the confinement mechanism of the model, the Feynman
diagrams do not contain branch points corresponding to on-shell quark 
production.

The values of the constituent quark masses $m_q$ and 
the infrared cut-off parameter $\lambda$ have been fixed from the 
analysis of a wide spectrum of data on meson and baryon decays with
\be
\def\arraystretch{1.5}
\begin{array}{ccccccc}
     m_{u,d}        &      m_s        &      m_c       & \lambda  &   
\\\hline
 \ \ 0.241\ \   &  \ \ 0.428\ \   &  \ \ 1.67\ \   & 
\ \ 0.181\ \   & \ {\rm GeV} 
\end{array}
\label{eq: fitmas}
\en
The values of the hadronic  size parameters are taken 
from our previous papers~\cite{Gutsche:2013pp,Gutsche:2013oea}:
\be
\def\arraystretch{1.5}
\begin{array}{cccc}
 \Lambda_{\Lambda_s}  &   \Lambda_{\Lambda_c} &   
\\\hline
   \ \ 0.492 \ \   &  \ \ 0.867\ \       & \ {\rm GeV} 
\end{array}
\label{eq: size}
\en

The results of our numerical two-loop calculation are well represented
by a double--pole parametrization
\be\label{eq:DPP}
F(q^2)=\frac{F(0)}{1 - a s + b s^2}\,, \quad s=\frac{q^2}{M_1^2} \,. 
\en 
For the $\Lambda_c \to \Lambda_s$ transition the parameters of the 
approximated form of the form factors
are given by
\be
\hspace*{-.5cm}
\begin{array}{ccccccc}
 &F_1^V & F_2^V & F_3^V 
 &F_1^A & F_2^A & F_3^A 
\\[1.5ex]
\hline\\[-2ex]
F(0) &  0.511  & 0.289 & -0.014 & 0.466 & -0.025 & -0.400
\\[1ex] 
a    &  1.736  & 1.970 &  5.053 & 1.594 & 0.321 &  2.083
\\[1ex] 
b    &  0.760  & 1.054 &  8.138 & 0.647 & 8.127 &  1.195
\\[1.1ex] 
\hline
\end{array}
\label{eq:ff_param}
\en

Let us take a closer look at the $q^2$ dependence of the
form factors $F_1^V(q^{2})\approx F_1^A(q^{2}) $. Their $q^2$ dependence is 
very close to a dipole behavior 
since one has $\sqrt{b} \sim a/2$ in both cases with a dipole 
mass $m_{\rm dipole} = M_1/\sqrt{a/2} \simeq 2.45$ GeV. The dipole mass is
quite close to the expected $(c\bar s)$ mass scale of $2.1121$ GeV set by the 
$D_s^\ast$ meson mass~\cite{Agashe:2014kda}. 

\section{Numerical results}

We shall present numerical results for the two cases $\ell^{+}=e^{+}$ 
and $\ell^{+}=\mu^{+}$. Note that the results for two modes are sometimes 
similar and sometimes not. In the calculations we use the baryon mass values 
$M_{\Lambda_c}=2.28646$ GeV and
$M_{\Lambda_s}=1.11568$ GeV~\cite{Agashe:2014kda}. 

\begin{figure}[htb]
\begin{center}
\vspace*{.2cm}
\epsfig{figure=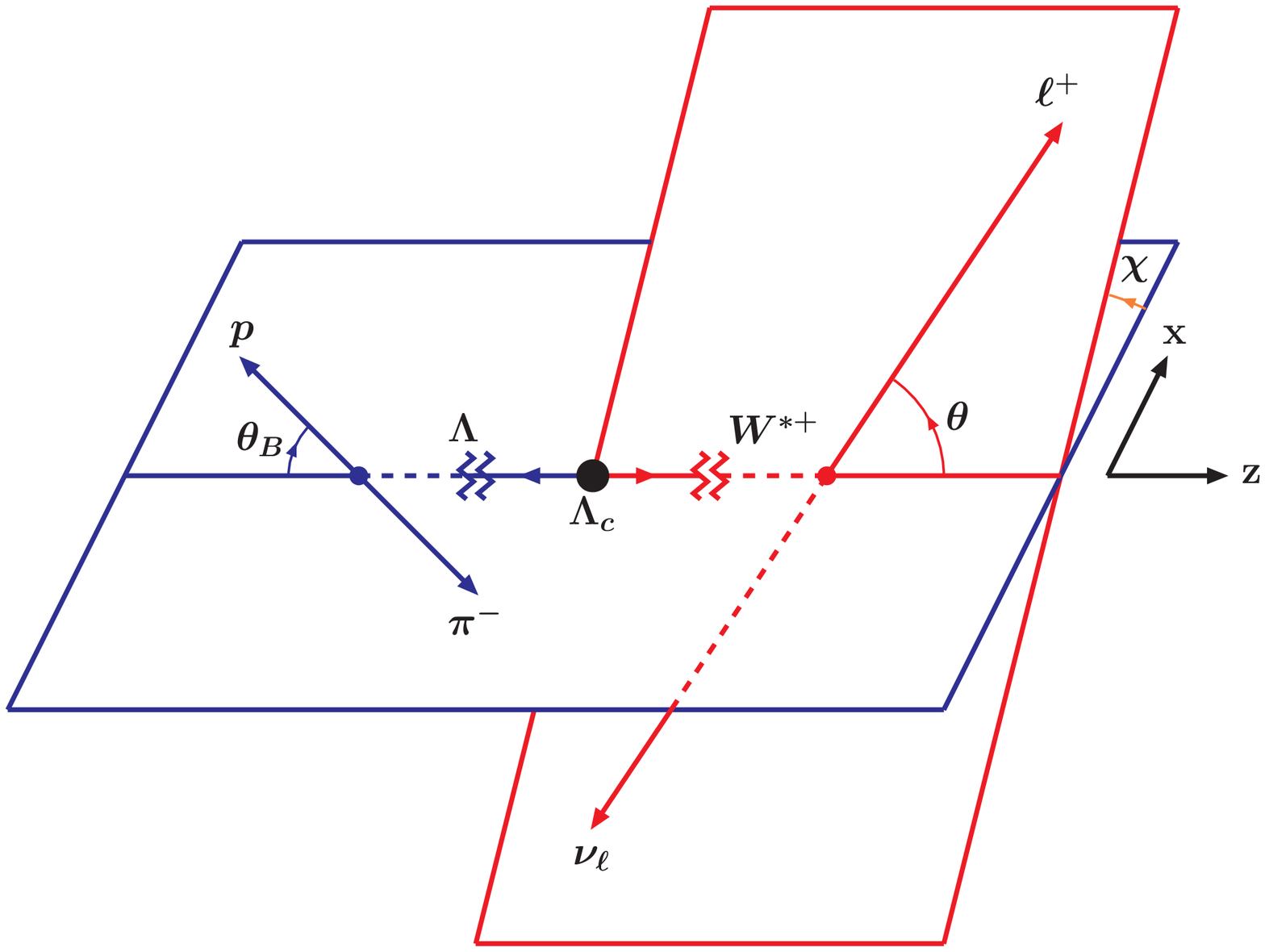,scale=.4}
\vspace*{-.75cm}
\caption{Definition of the polar 
and the azimuthal angles.} 
\label{fig:angles}
\end{center}

\begin{center}
\vspace*{.2cm}
\epsfig{figure=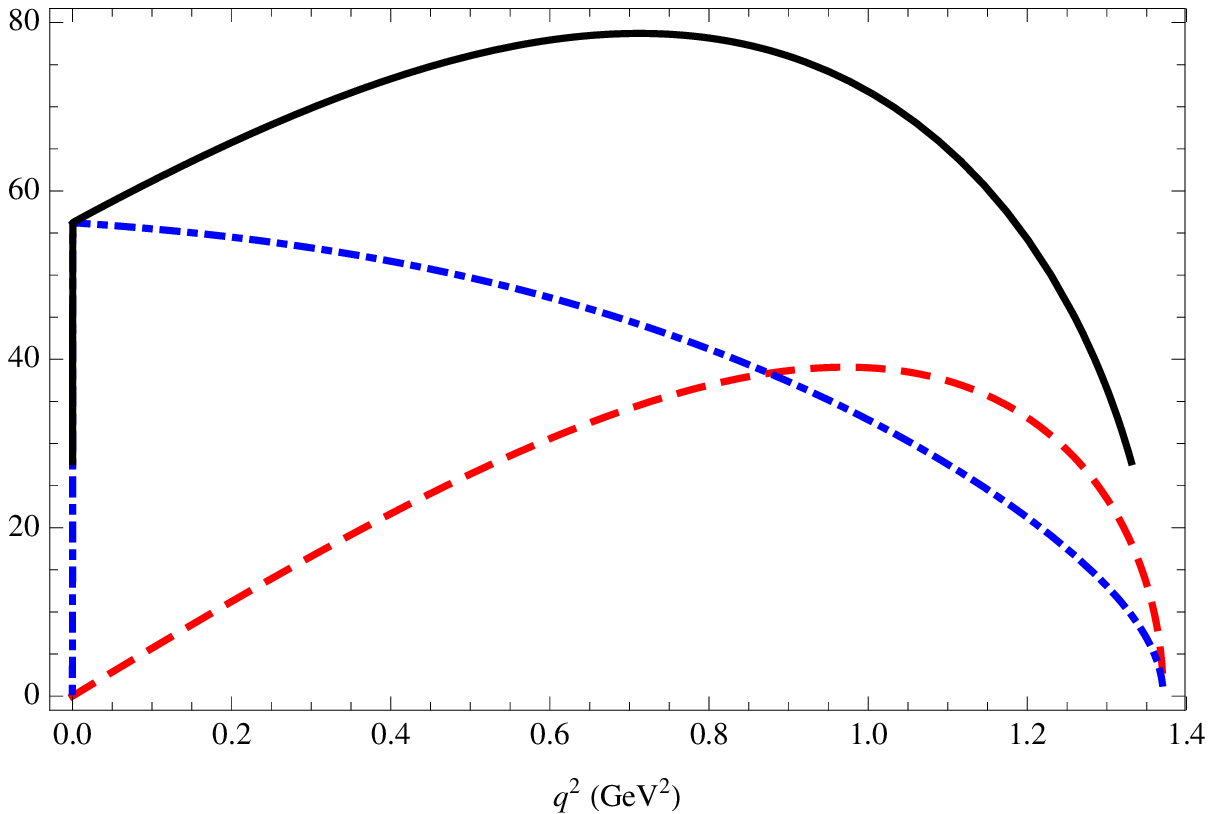,scale=.45}
\caption{The  $q^{2}$-dependence of the partial rates
$d\Gamma_{U}/dq^{2}$ (dashed), $d\Gamma_{L}/dq^{2}$ (dot-dashed) and their 
sum $d\Gamma_{U+L}/dq^{2}$ (solid)
for the $e^+$-mode (in units of $10^{-15}$~GeV$^{-1}$).
\label{fig:dUL}
}
\end{center}

\begin{center}
\vspace*{.2cm}
\epsfig{figure=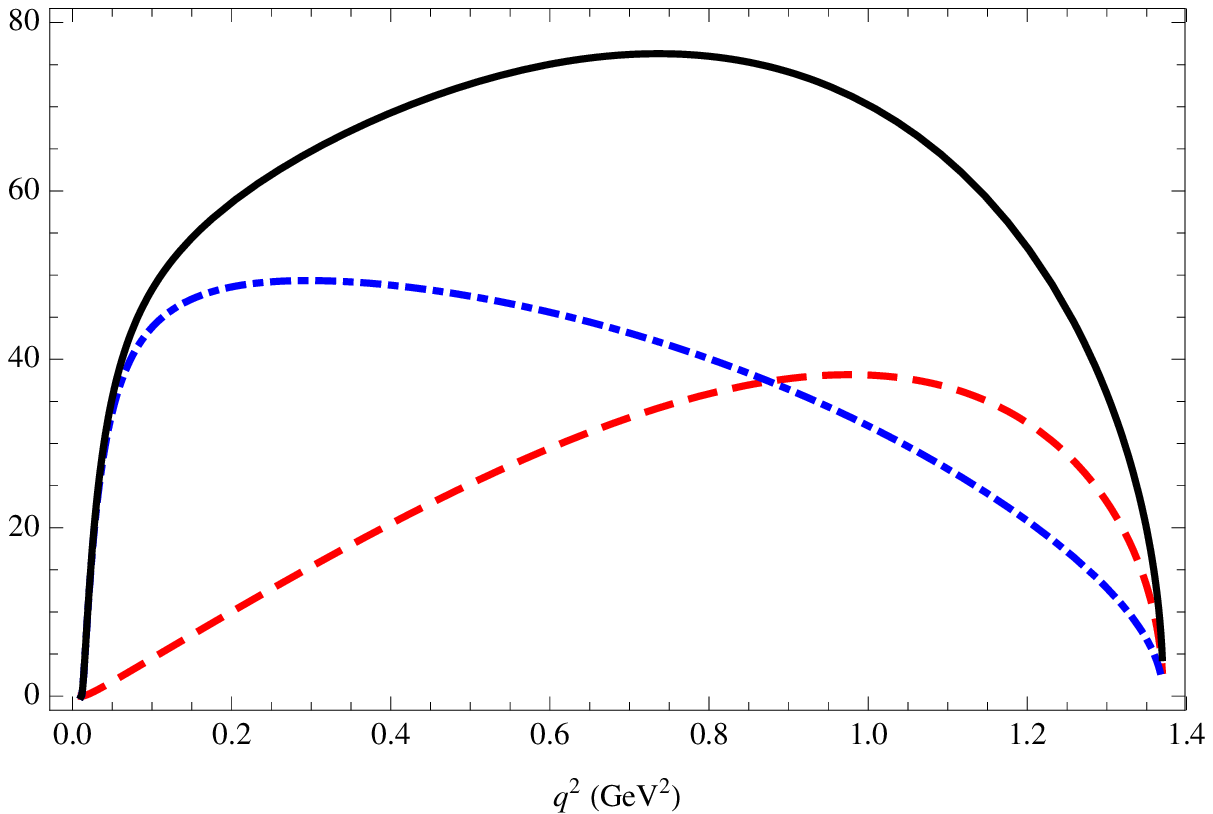,scale=.45}
\caption{The  $q^{2}$-dependence of the partial rates
$d\Gamma_{U}/dq^{2}$ (dashed), $d\Gamma_{L}/dq^{2}$ (dot-dashed) and their 
sum $d\Gamma_{U+L}/dq^{2}$ (solid)
for the $\mu^+$-mode (in units of $10^{-15}$~GeV$^{-1}$).
\label{fig:dULmu}
}
\end{center}
\end{figure}

In Figs.~\ref{fig:dUL} and~\ref{fig:dULmu} 
we display the $q^{2}$--dependence of the
partial differential rates $d\Gamma_{U}/dq^{2}$, $d\Gamma_{L}/dq^{2}$ and 
the total differential rate $d\Gamma_{U+L}/dq^{2}$ for the $e$- and $\mu$-mode.
The transverse rate dominates in the low recoil region while the longitudinal 
rate dominates in the large recoil region. In case of the $e$-mode the 
longitudinal and thereby the total rate shows a step-like behavior close 
to the threshold $q^{2}=m^{2}_{e}$.

In Fig.~4 we show the $q^{2}$--dependence of the lepton-side 
forward-backward asymmetry $A^{\ell}_{FB}(q^{2})$,  
which is negative and very similar for both $e$- and 
$\mu$-mode at $q^2 \ge 1$~GeV$^2$.  
At zero recoil $A^{\ell}_{FB}(q^{2})$ goes to zero due to 
the zero recoil relations 
${\cal H}_{P}={\cal H}_{SL}=0$. In the large recoil limit $A^{e}_{FB}(e)$ 
goes to zero due to the longitudinal dominance in the partial rates. 

In Fig.~5 we display the $q^{2}$--dependence of the 
convexity parameter $C_{F}$, which is similar for both modes 
for $q^2~\ge~0.6$~GeV$^2$ and different for $q^2 < 0.6$~GeV$^2$.  
At zero recoil 
$C_{F}$ goes to zero for both modes
due to the zero recoil relation ${\cal H}_{U}=2{\cal H}_{L}$. 
For the $e$--mode one finds $C_{F} \to -1.5$ at maximal recoil due to the 
longitudinal dominance while $C_{F} \to 0$ for the $\mu$--mode at maximal 
recoil $q^{2}=m^{2}_{\mu}$ due to the overall factor $(1-2\delta)$ 
in~(\ref{eq:CF}). In both modes $C_{F}$ is 
large and negative which implies that the $\cos\theta$--distribution is
strongly parabolic in terms of a downward open tilted parabola.  

In Figs.~6 and~7 we show the longitudinal and 
trans\-ver\-se polarization components of the $\Lambda$ defined in 
Eq.~(\ref{eq:Pol_had}), which are similar 
for $e$- and $\mu$-modes in case of both polarizations. 
The magnitude of the $\Lambda$ polarization shown in 
Fig.~8 is also similar for both modes and is quite large. 

In Figs.~9 and~10 we show the $q^{2}$--dependence of the longitudinal and 
transverse polarization components of the charged lepton. In the case of 
the electron the two curves reflect the chiral limit of a massless lepton 
in which the lepton is purely left-handed. The behavior of the two
polarization components in the $\mu$--mode is distinctly different. 
The longitudinal polarization is reduced from $1$ to 
a negative number of $-0.34$ at zero recoil, 
while the transverse polarization can become 
negative and quite large on magnitude towards maximal recoil. 
At zero recoil the transverse polarization of the charged lepton tends 
to zero in agreement with the vanishing of 
${\cal H}_{P}$ and ${\cal H}_{SL}$ at zero recoil. 
The total polarization of the lepton shown in Fig.~11 is 
maximal in the $e$--mode and somewhat
reduced but still quite large in the $\mu$--mode. 

In Table~\ref{Br_LcLs} 
we present our predictions for the semileptonic branching ratios 
of the $\Lambda_c$ in \%  and compare them with data from the 
Belle~\cite{Zupanc:2013iki} and BESIII~\cite{Ablikim:2015prg} Collaborations.  
We have used the value for the $\Lambda_c$--lifetime 
from the Particle Data Group~\cite{Agashe:2014kda} 
$\tau_{\Lambda_c} = (2.0\pm 0.06) \times 10^{-13}\,$s. 
One can see that our results are in a good agreement with Belle data and 
close to the lower value of the BESIII result. 
In Table~\ref{LcL_comparison} we compare our predictions with 
previous theoretical results~\cite{Ivanov:1996fj},\cite{Buras:1976dg}-\cite{Liu:2009sn} 
for $\Lambda_c^+ \to \Lambda^0 e^+ \nu_e$. 
For some approaches in brackets we indicate 
the result by taking into account the $SU(6)$ spin-flavor suppression factor 
equal to $1/3$ (see detailed discussion in Ref.~\cite{Korner:1994nh}).  
$q^{2}$--averages of helicity structure functions in units of 
$10^{-15}$~GeV are listed in Table~\ref{tab:bilinears-numerics}. 
We do not display helicity flip results for the $e^+ \nu_e$ mode 
because they are of order $10^{-6}-10^{-7}$ in the above units. 
Partial rates, from which we can compile
the total rate, are listed in Table~\ref{tab:part-rates}, 
Again we list the partial and total rates in units of 
$10^{-15}$~GeV. 
The numbers show that the partial flip rates make up $34.2\,\%$ of the total 
rate where the biggest contribution comes from the scalar rate with 
$20.4\,\%$. The results for the mean values of the asymmetry parameters 
are shown in Table~\ref{tab:asymmetry-results}. 
When calculating the $q^{2}$--averages one has to 
remember to include the $q^{2}$--dependent factor 
$(q^2-m_\ell^2)^2 |\mathbf{p_2}|/q^2$ in the numerator and denominator of the
relevant asymmetry expressions. In 
most of the shown cases the mean values change considerably when going from 
the $e$-- to the $\mu$--modes
including even a sign change in $\la A_{FB}^{\ell} \ra$. 

\begin{widetext}
\begin{table}[htb]
\caption{Semileptonic branching ratios
of the $\Lambda_c$ in \%.}
\def\arraystretch{1}
\begin{center}
\vspace*{-.1cm}
\begin{tabular}{cll}
\hline
Mode  & Our results & 
Data  \\[1.5ex]
\hline
$\Lambda_c^+ \to \Lambda^0 e^+ \nu_e$  
& \ \ \ 2.78   & $(2.9 \pm 0.5)$ \ Belle            \\
&        & $(3.63 \pm 0.38 \pm 0.20)$ \ BESIII  \\[1.5ex]
$\Lambda_c^+ \to \Lambda^0 \mu^+ \nu_\mu$  
& \ \ \ 2.69   & $(2.7 \pm 0.6)$ \ Belle            \\
\hline 
\end{tabular} 
\label{Br_LcLs} 
\end{center} 

\begin{center}
\caption{Comparison of our predictions with 
previous theoretical results for 
Br$(\Lambda_c^+ \to \Lambda^0 e^+ \nu_e)$ in \%.}
\vspace*{-.1cm}
\begin{tabular}{cccccccccccc}
\hline 
Our &
Ref.~\cite{Buras:1976dg}             & 
Ref.~\cite{Gavela:1979wk}            & 
Ref.~\cite{PerezMarcial:1989yh}      & 
Ref.~\cite{Singleton:1990ye}         & 
Ref.~\cite{Hussain:1990ai}           & 
Ref.~\cite{Cheng:1995fe}             & 
Ref.~\cite{Datta:1995mv}             & 
Ref.~\cite{Ivanov:1996fj}            & 
Ref.~\cite{Luo:1998wg}               & 
Ref.~\cite{MarquesdeCarvalho:1999ia} &
Ref.~\cite{Liu:2009sn} \\
\hline 
2.78   & 12 (4) & 3 (1) & 3.4 (1.12) & 2 & 4.4 (1.46) 
& 1.42 & 1.07   & 1.44  & 1.4        & 2.64 $\pm$ 0.36 & 3.05 $\pm$ 0.27 \\  
       &        &       & 2.6 (0.86) &   &  
&      &        &       &            &               & 1.96 $\pm$ 0.32 \\       
\hline\\
\end{tabular}
\label{LcL_comparison} 
\end{center} 

\vspace*{-1.1cm}
\begin{center}
\caption{$q^{2}$--averages of helicity structure functions in units of 
$10^{-15}$~GeV.}  
\def\arraystretch{1}
\vspace*{-.2cm}
\begin{tabular}{c|ccccccccccc}
\hline
 \qquad \qquad 
&\quad $\Gamma_{U}$    \qquad   
&\quad $\Gamma_{L}$    \qquad 
&\quad $\phantom{S}$    \qquad   
&\quad $\Gamma_{LT}$   \qquad   
&\quad $\phantom{SL}$   \qquad  
&\quad $\Gamma_{SL}$   \qquad   
&\quad $\Gamma_{P}$    \qquad    
&\quad $\Gamma_{L_{P}}$   \qquad     
&\quad $\Gamma_{LT_{P}}$   \qquad 
&\quad $\phantom{ST_P}$ \qquad 
&\quad $\phantom{SL_P}$ \qquad 
\\
\hline
\qquad $e^+ \nu_e$\qquad  \quad   
&\quad $ 35.6 $  \qquad  
&\quad $ 55.8 $  \qquad 
&\quad $ \phantom{58.5} $  \qquad  
&\quad $-17.7 $  \qquad 
&\quad $ \phantom{53.9 }$ \qquad  
&\quad $ 53.9 $  \qquad  
&\quad $-25.7 $  \qquad 
&\quad $-53.6 $  \qquad  
&\quad $-8.8 $  \qquad  
&\quad $\phantom{-21.5} $  \qquad  
&\quad $\phantom{-56.9} $  \qquad 
\\
\qquad $\mu^+ \nu_\mu$ \qquad \quad   
&\quad $ 34.3 $ \qquad  
&\quad $ 50.3 $ \qquad 
&\quad $ \phantom{52.9 }$ \qquad  
&\quad $-17.1 $ \qquad 
&\quad $ \phantom{48.5 }$ \qquad  
&\quad $ 48.5 $ \qquad  
&\quad $-24.7 $ \qquad 
&\quad $-48.1 $ \qquad  
&\quad $-8.3 $ \qquad  
&\quad $\phantom{-20.6} $ \qquad  
&\quad $\phantom{-51.4} $ \qquad 
\\ 
\hline
 \qquad\qquad 
&\quad $\widetilde{\Gamma}_{U}$    \qquad 
&\quad $\widetilde{\Gamma}_{L}$    \qquad 
&\quad $\widetilde{\Gamma}_{S}$    \qquad  
&\quad $\widetilde{\Gamma}_{LT}$   \qquad
&\quad $\widetilde{\Gamma}_{S_{P}}$   \qquad   
&\quad $\widetilde{\Gamma}_{SL}$   \qquad 
&\quad $\widetilde{\Gamma}_{P}$    \qquad    
&\quad $\widetilde{\Gamma}_{L_P}$   \qquad 
&\quad $\widetilde{\Gamma}_{LT_P}$   \qquad 
&\quad $\widetilde{\Gamma}_{ST_P}$ \qquad 
&\quad $\widetilde{\Gamma}_{SL_P}$ \qquad 
\\
\hline
\qquad $\mu^+ \nu_\mu$ \qquad \quad   
&\quad $ 0.31 $ \qquad  
&\quad $ 0.91 $ \qquad 
&\quad $ 0.93 $ \qquad  
&\quad $-0.15 $ \qquad
&\quad $-0.90 $ \qquad 
&\quad $ 0.91 $ \qquad  
&\quad $-0.24 $ \qquad 
&\quad $-0.90 $ \qquad  
&\quad $-0.10 $ \qquad  
&\quad $-0.19 $ \qquad  
&\quad $-0.92 $ \qquad 
\\
\hline
\end{tabular}
\label{tab:bilinears-numerics}
\end{center}

\caption{Partial rates contributing to the 
total rates $\Gamma(\Lambda_c^+ \to \Lambda^0 + \ell^+ \nu_\ell)$ 
in units of $10^{-15}$~GeV.} 
\begin{center}
\vspace*{-.2cm}
\begin{tabular}{c|ccccc|c}
\hline
   & \quad$\Gamma_U$\quad & \quad$\Gamma_L$\quad & \quad$\widetilde\Gamma_{U}$\quad
&\quad$\widetilde\Gamma_{L}$\quad&
\quad$\widetilde\Gamma_{S}$\quad
& \quad$\Gamma_{\rm tot}$\quad \\[1.1ex]
\hline\\[-3ex]
$e^+\nu_e$       &  35.6  &  55.8  & - & - & -       & 91.4   \\[1.1ex]
$\mu^+\nu_\mu$   &  34.3  &  50.3  &0.3&0.9& 2.8  & 88.6   \\[1.1ex]
\hline
\end{tabular}
\label{tab:part-rates}
\end{center}

\caption{Mean values of the asymmetry parameters.} 
\begin{center}
\vspace*{-.4cm}
\begin{tabular}{c|ccccccc}
\hline
       & \quad $<A_{FB}^{\ell}>$ \quad  & 
\quad $\la C_F \ra $\quad  &  \quad $\la P^h_z \ra$ \quad  & \quad $\la P^h_x \ra$ \quad 
 &  \quad $\la P^\ell_z \ra$ \quad  & \quad $\la P^\ell_x \ra$ \quad & \quad $\la \gamma \ra$ \qquad
\\[1.2ex]
\hline\\[-2ex]
$e^+\nu_e$        &  
-0.21  &  -0.62 & -0.87 & -0.32 & 1.00 & -0.001 & -0.25  
\\[1.2ex] 
$\mu^+\nu_\mu$  &  
-0.24  &  -0.54 & -0.87 & -0.33 & 0.91 & -0.18  & -0.26 
\\[1.2ex] 
\hline
\end{tabular}
\label{tab:asymmetry-results}
\end{center}
\end{table}
\end{widetext}

\section{Summary and conclusions}

Let us summarize the main results of our paper. 
We have used the helicity formalism to study the angular decay 
distribution in the semileptonic decay 
$\Lambda_c^+ \to \Lambda^{0} + \ell^+ \nu_\ell$ as well as the 
corresponding cascade decay 
$\Lambda_c^+ \to \Lambda^{0} (\to p + \pi^-) + \ell^+ \nu_\ell$. 
Starting from the angular decay distribution we have defined a number of 
polarization observables for which we have provided numerical results 
using form factor results from the covariant confined quark model. 
Our predictions for the absolute branching fractions for $\Lambda_c^+ \to \Lambda \ell^+ \nu_\ell$
decays are in agreement with the central value of data 
from Belle~\cite{Zupanc:2013iki} and with the lower limit of 
the data from the BESIII Collaboration~\cite{Ablikim:2015prg}. 

\begin{acknowledgments}

This work was supported by Tomsk State University Competitiveness 
Improvement Program and the Russian Federation program ``Nauka'' 
(Contract No. 0.1526.2015, 3854). 
M.A.I.\ acknowledges the support from Mainz Institute for Theoretical 
Physics (MITP). M.A.I. and J.G.K. thank the Heisenberg-Landau Grant for
support.  

\end{acknowledgments}



\newpage 

\begin{figure}[ht]
\begin{center}
\hspace*{-0.5cm}
\begin{tabular}{lr}
\includegraphics[width=0.4\textwidth]{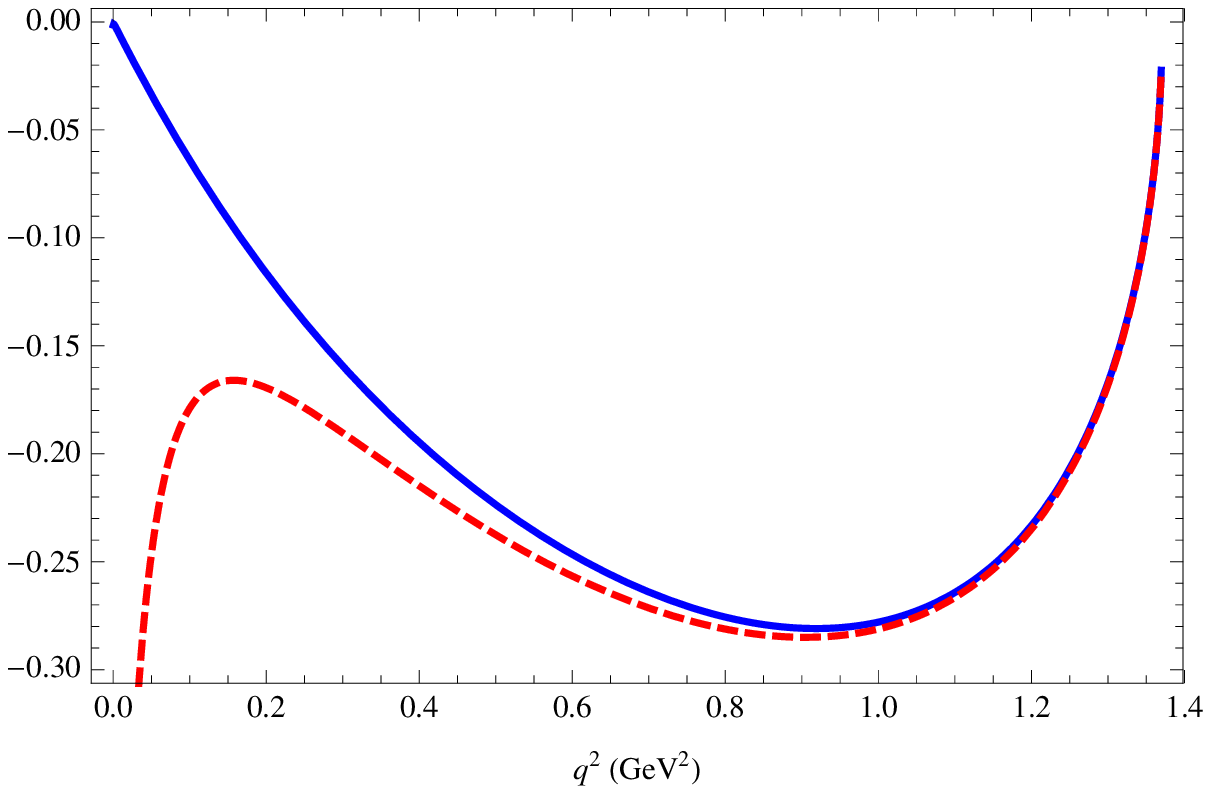}   & \hspace*{.5cm}
\includegraphics[width=0.4\textwidth]{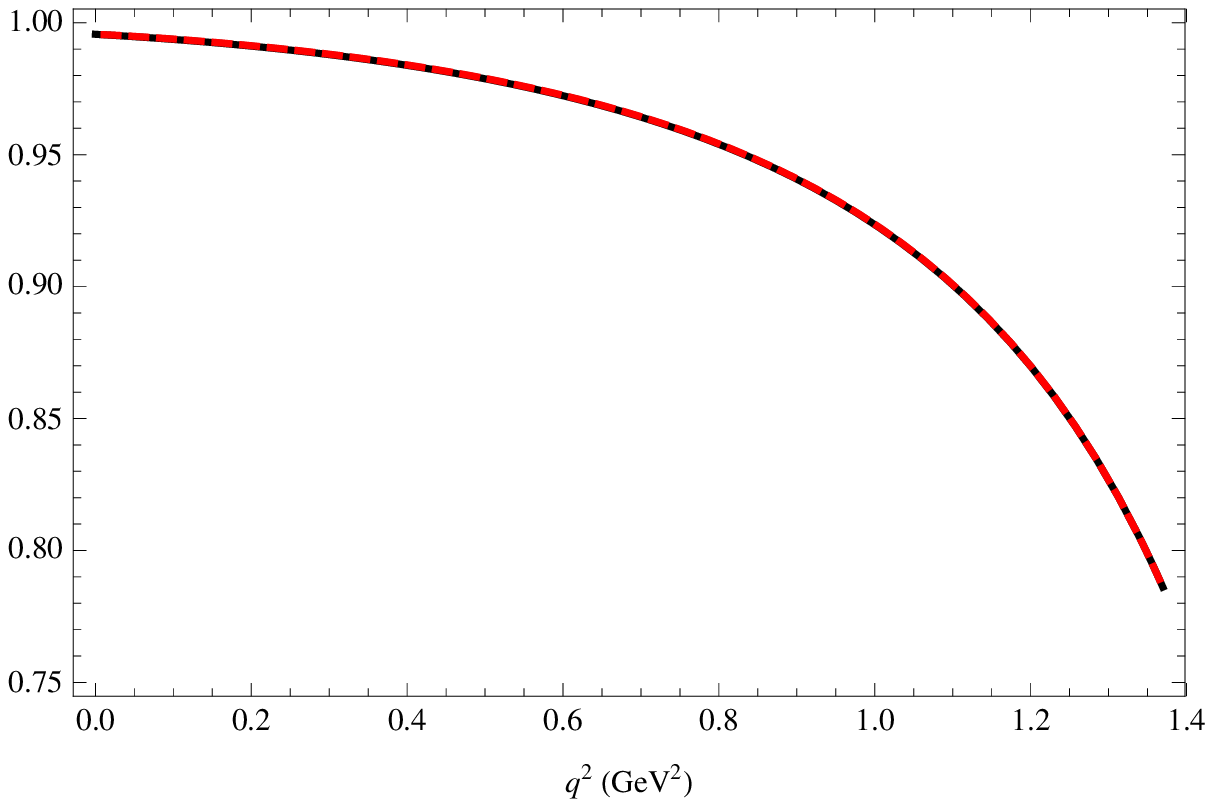}      \\
\hspace*{0.5cm} 
FIG. 4 $A^{\ell}_{FB}(q^{2})$ plot: $e^+$ (solid) 
and $\mu^+$ (dashed) & 
FIG. 8 $|\vec {P}\,^h|(q^{2})$ plot: $e^+$ (solid) 
and $\mu^+$ (dashed).\\[2ex]
\includegraphics[width=0.4\textwidth]{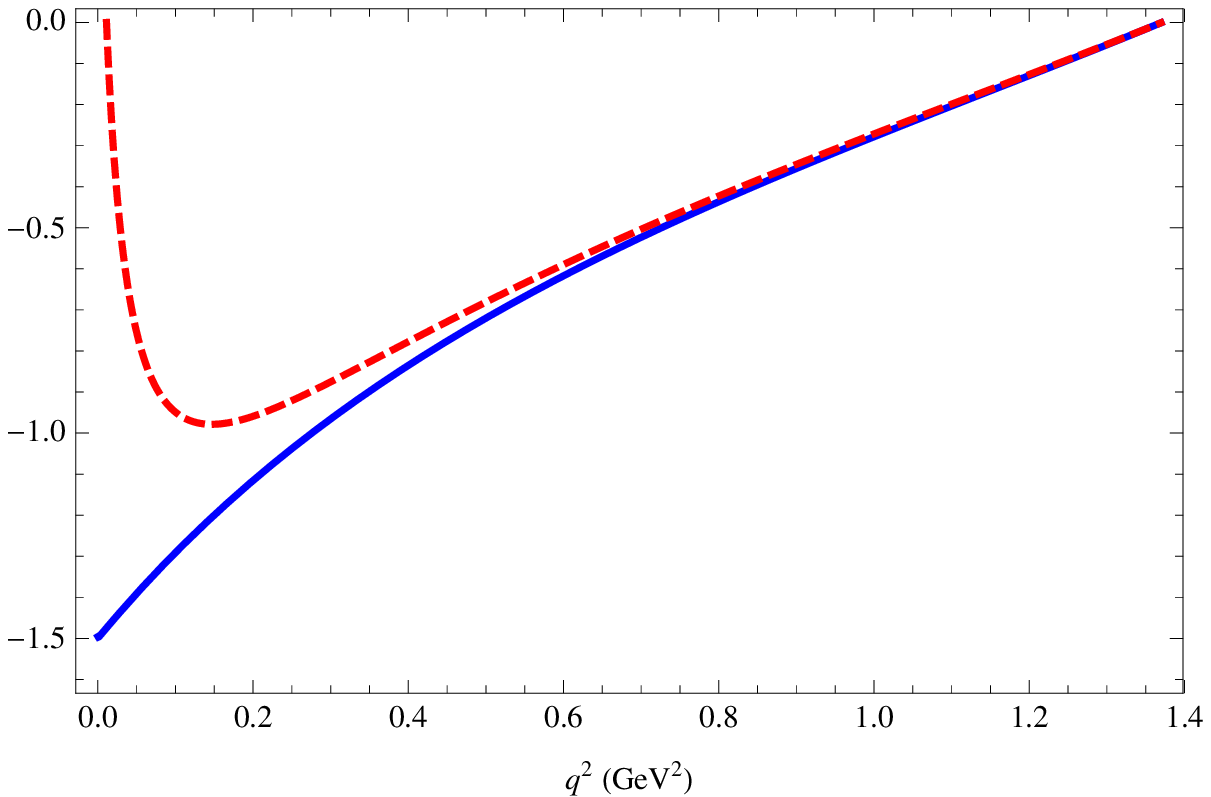}   & \hspace*{.5cm}
\includegraphics[width=0.4\textwidth]{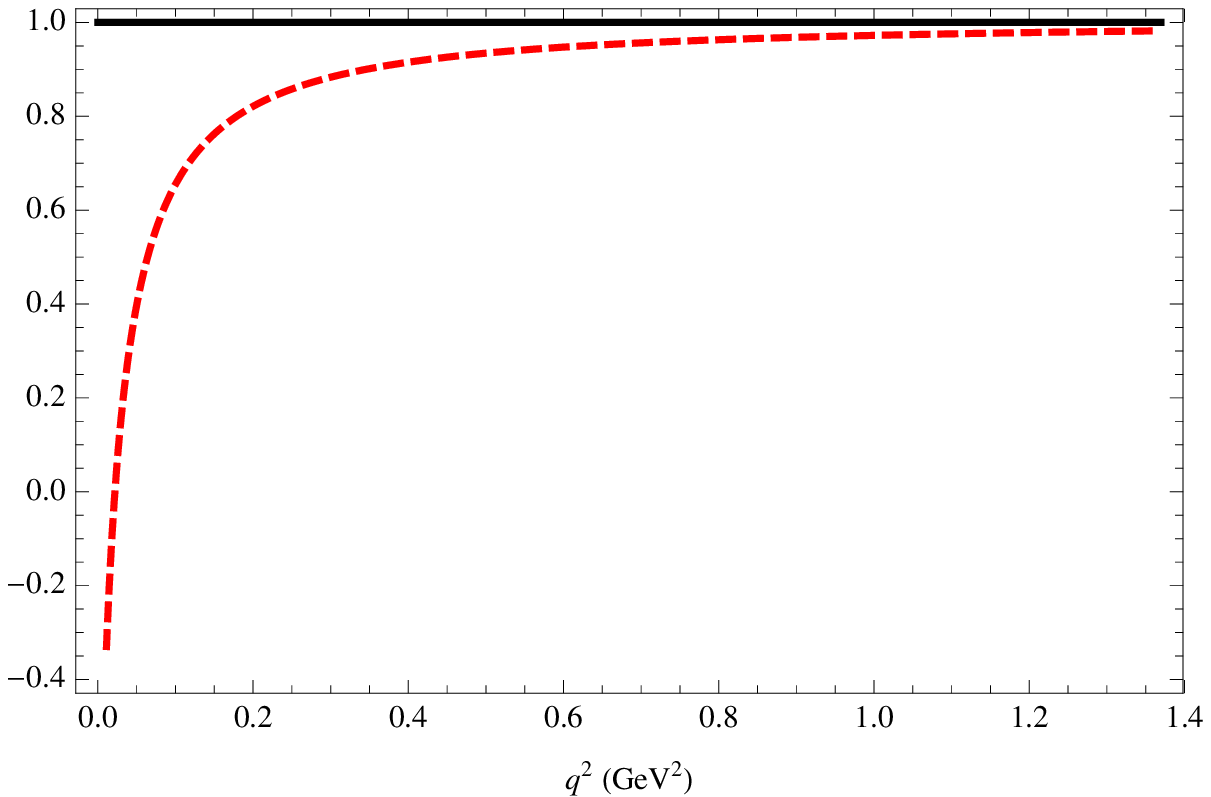}      \\
\hspace*{0.5cm} FIG. 5 $C_{F}(q^{2})$ plot: $e^+$ (solid) 
and $\mu^+$ (dashed) & 
FIG. 9 $P^\ell_z(q^{2})$ plot: $e^+$ (solid) 
and $\mu^+$ (dashed).\\[2ex]
\includegraphics[width=0.4\textwidth]{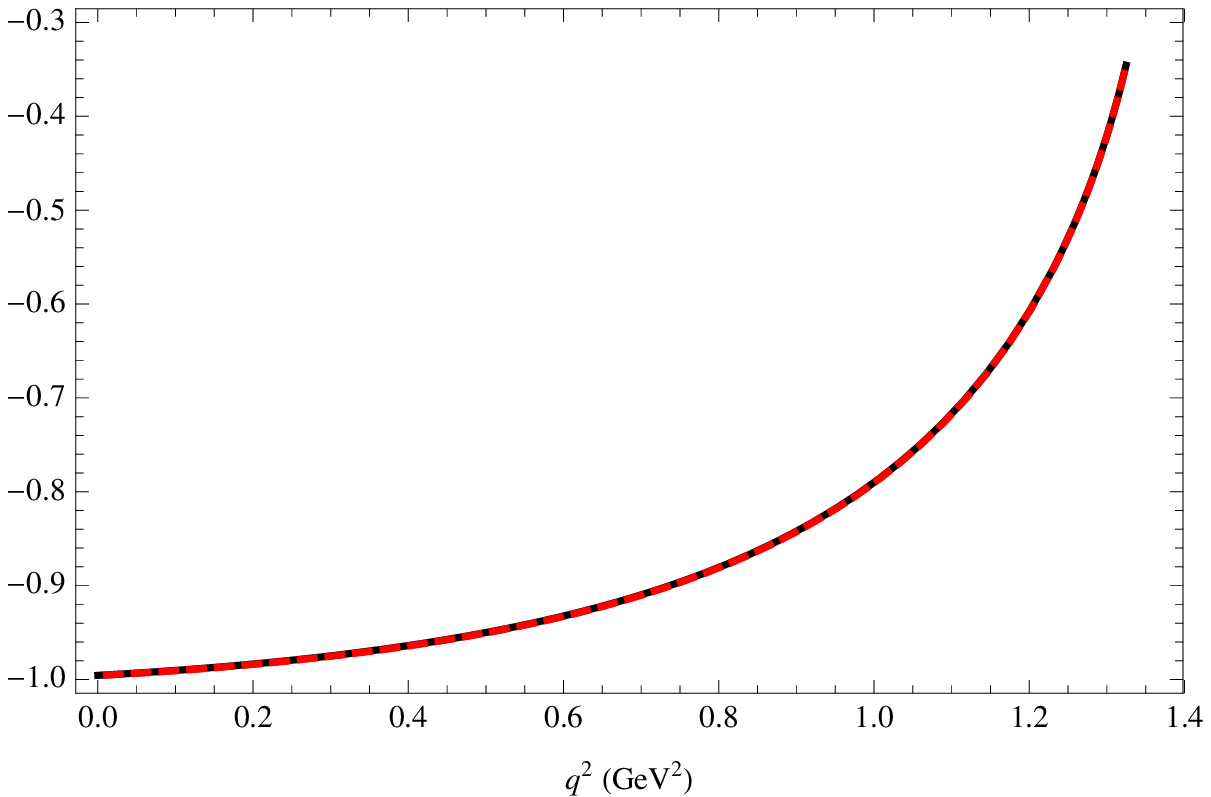}   & \hspace*{.5cm}
\includegraphics[width=0.4\textwidth]{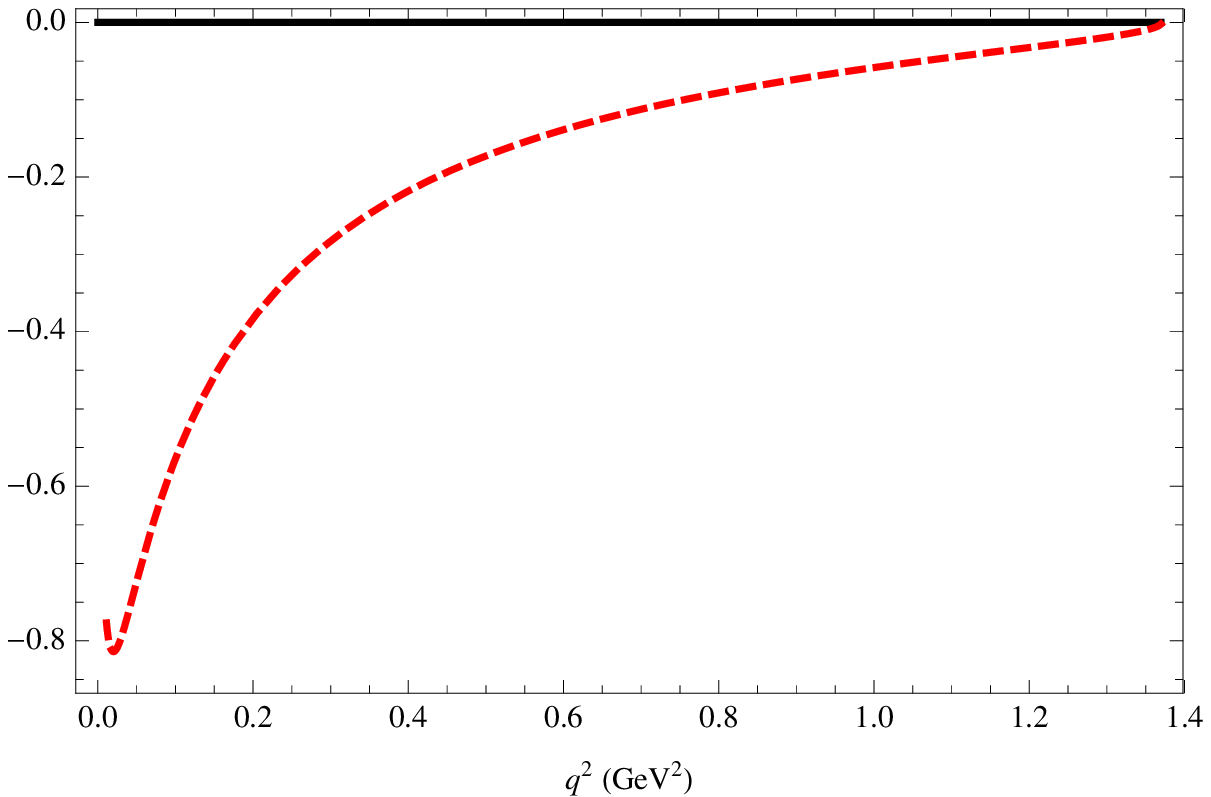}      \\
\hspace*{0.5cm} FIG. 6 $P^h_z(q^{2})$ plot: $e^+$ (solid) 
and $\mu^+$ (dashed) & 
FIG. 10 $P^\ell_x(q^{2})$ plot: $e^+$ (solid) 
and $\mu^+$ (dashed).\\[2ex]
\includegraphics[width=0.4\textwidth]{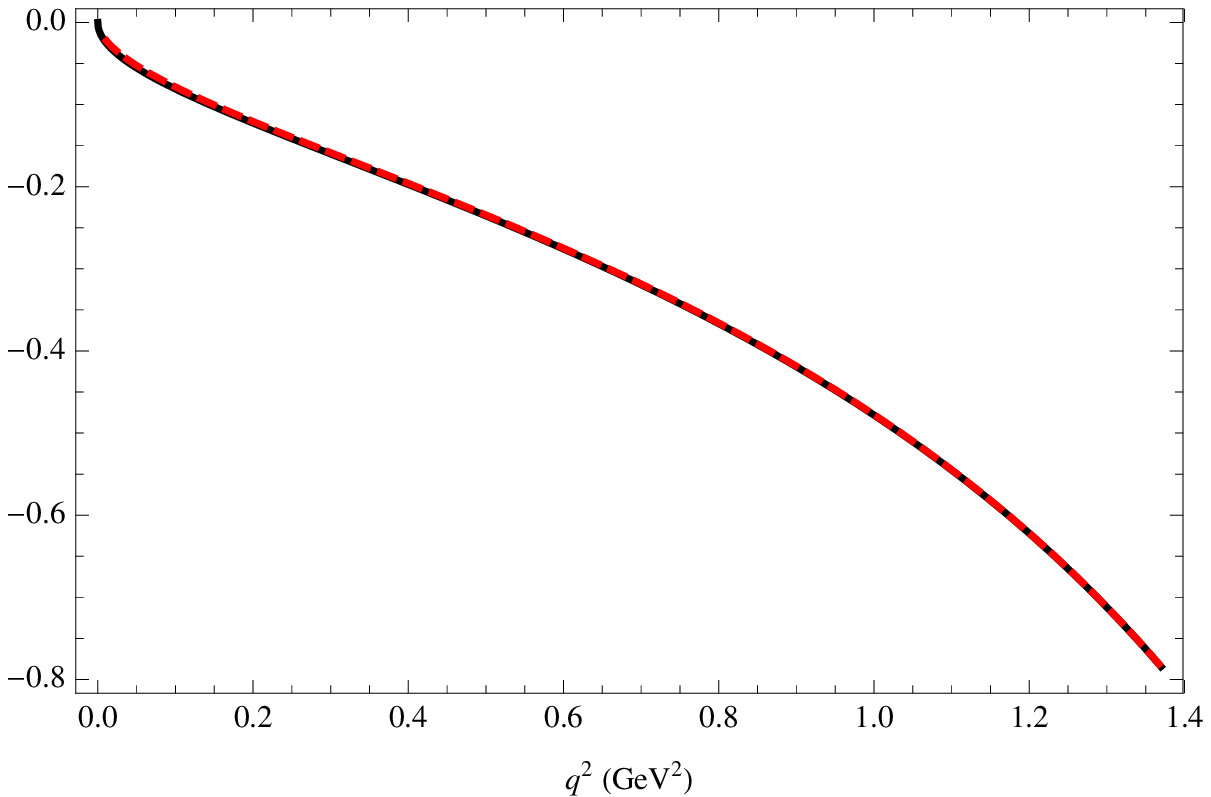}   & \hspace*{.5cm}
\includegraphics[width=0.4\textwidth]{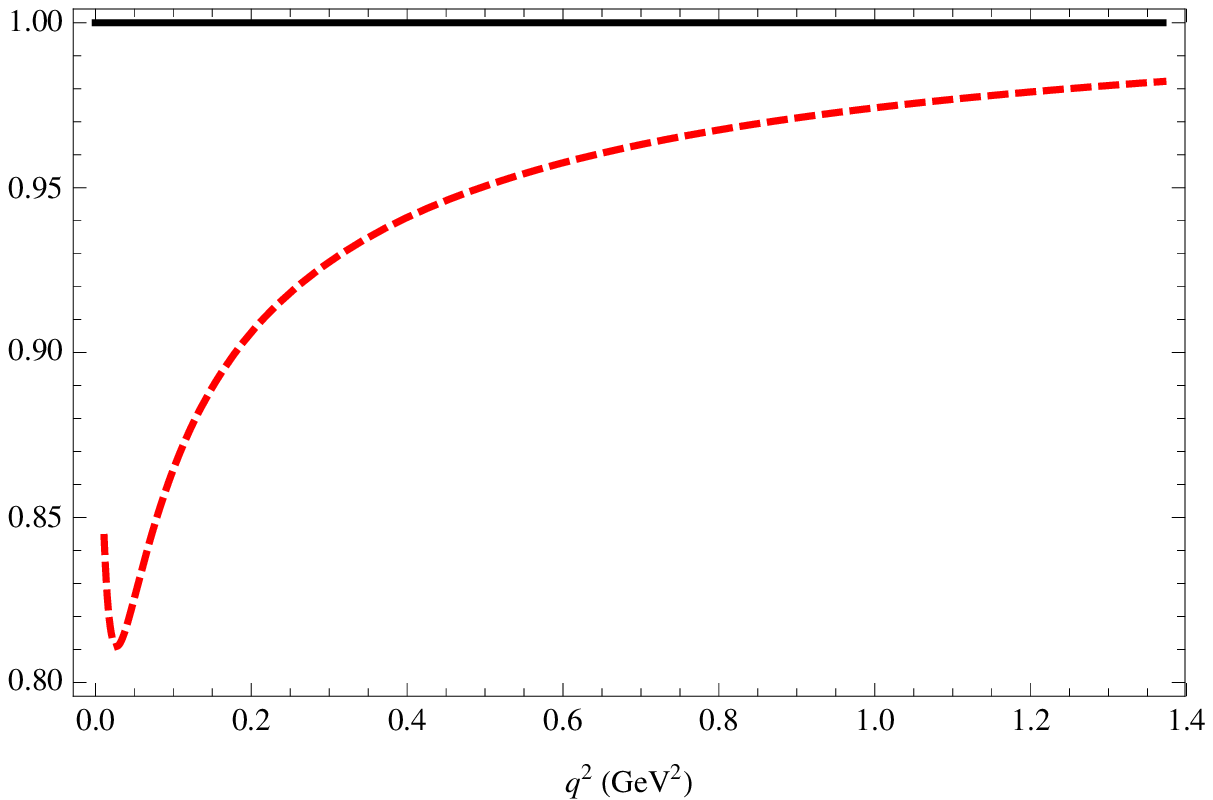}      \\
\hspace*{0.5cm} FIG. 7 $P^h_x(q^{2})$ plot: $e^+$ (solid) 
and $\mu^+$ (dashed) & 
FIG. 11 $|\vec{P}\,^\ell|(q^{2})$ plot: $e^+$ (solid) 
and $\mu^+$ (dashed).\\[2ex]
\end{tabular}
\end{center}
\end{figure}


\begin{thebibliography}{99}

\bibitem{Zupanc:2013iki} 
  A.~Zupanc {\it et al.} (Belle Collaboration),
  Phys.\ Rev.\ Lett.\  {\bf 113}, no. 4, 042002 (2014)
  [arXiv:1312.7826 [hep-ex]].

\bibitem{Agashe:2014kda} 
  K.~A.~Olive {\it et al.}  (Particle Data Group Collaboration),
  Chin.\ Phys.\ C {\bf 38}, 090001 (2014).

\bibitem{Ablikim:2015prg} 
  M.~Ablikim {\it et al.} (BESIII Collaboration),
  Phys.\ Rev.\ Lett.\  {\bf 115}, no. 22, 221805 (2015)    
  [arXiv:1510.02610 [hep-ex]].
 

\bibitem{Ivanov:1996fj}
  M.~A.~Ivanov, V.~E.~Lyubovitskij, J.~G.~K\"orner and P.~Kroll,
  Phys.\ Rev.\ D {\bf 56}, 348 (1997) 
  [hep-ph/9612463].

\bibitem{Gutsche:2012ze} 
T.~Gutsche, M.~A.~Ivanov, J.~G.~K\"orner, V.~E.~Lyubovitskij and P.~Santorelli,
   Phys.\ Rev.\ D {\bf 86}, 074013 (2012)   
  [arXiv:1207.7052 [hep-ph]].

\bibitem{Gutsche:2013pp} 
T.~Gutsche, M.~A.~Ivanov, J.~G.~K\"orner, V.~E.~Lyubovitskij and P.~Santorelli,
  Phys.\ Rev.\ D {\bf 87}, 074031 (2013)  
  [arXiv:1301.3737 [hep-ph]].

\bibitem{Gutsche:2013oea}
  T.~Gutsche, M.~A.~Ivanov, J.~G.~K\"orner, V.~E.~Lyubovitskij 
  and P.~Santorelli,
  Phys.\ Rev.\ D {\bf 88}, 114018 (2013) 
  [arXiv:1309.7879 [hep-ph]].

\bibitem{Gutsche:2014zna} 
  T.~Gutsche, M.~A.~Ivanov, J.~G.~K\"orner, 
  V.~E.~Lyubovitskij and P.~Santorelli,
  Phys.\ Rev.\ D {\bf 90}, no. 11, 114033 (2014)
  [arXiv:1410.6043 [hep-ph]].

\bibitem{Gutsche:2015mxa} 
  T.~Gutsche, M.~A.~Ivanov, J.~G.~K\"orner, V.~E.~Lyubovitskij, 
  P.~Santorelli and N.~Habyl,
  Phys.\ Rev.\ D {\bf 91}, no. 7, 074001 (2015)
  [Phys.\ Rev.\ D {\bf 91}, no. 11, 119907 (2015)]
  [arXiv:1502.04864 [hep-ph]]; 
  N.~Habyl, T.~Gutsche, \\{} M.~A.~Ivanov, J.~G.~K\"orner, 
  V.~E.~Lyubovitskij and \\{} P.~Santorelli,
  Int.\ J.\ Mod.\ Phys.\ Conf.\ Ser.\  {\bf 39}, 1560112 (2015) 
  [arXiv:1509.07688 [hep-ph]].

\bibitem{Korner:1989ve} 
  J.~G.~K\"orner and G.~A.~Schuler,
  Phys.\ Lett.\ B {\bf 231}, 306 (1989).

\bibitem{Korner:1989qb} 
  J.~G.~K\"orner and G.~A.~Schuler,
  Z.\ Phys.\ C {\bf 46}, 93 (1990).

\bibitem{Kruger:2005ep}
  F.~Kr\"uger and J.~Matias,
  Phys.\ Rev.\ D {\bf 71}, 094009 (2005). 

\bibitem{Korner:1994nh} 
  J.~G.~Korner, M.~Kramer and D.~Pirjol,
  Prog.\ Part.\ Nucl.\ Phys.\  {\bf 33}, 787 (1994) 
  [hep-ph/9406359].

\bibitem{Buras:1976dg} 
  A.~J.~Buras,
  Nucl.\ Phys.\ B {\bf 109}, 373 (1976).

\bibitem{Gavela:1979wk} 
  M.~B.~Gavela,
  Phys.\ Lett.\ B {\bf 83}, 367 (1979).

\bibitem{PerezMarcial:1989yh} 
  R.~Perez-Marcial, R.~Huerta, A.~Garcia and M.~Avila-Aoki,
  Phys.\ Rev.\ D {\bf 40}, 2955 (1989)
  [Phys.\ Rev.\ D {\bf 44}, 2203 (1991)].

\bibitem{Singleton:1990ye} 
  R.~L.~Singleton,
  Phys.\ Rev.\ D {\bf 43}, 2939 (1991).

\bibitem{Hussain:1990ai} 
  F.~Hussain and J.~G.~Korner,
  Z.\ Phys.\ C {\bf 51}, 607 (1991).

\bibitem{Cheng:1995fe} 
  H.~Y.~Cheng and B.~Tseng,
  Phys.\ Rev.\ D {\bf 53}, 1457 (1996)
  [Phys.\ Rev.\ D {\bf 55}, 1697 (1997)]. 

\bibitem{Datta:1995mv} 
  A.~Datta,
  hep-ph/9504429.

\bibitem{Luo:1998wg} 
  C.~W.~Luo,
  Eur.\ Phys.\ J.\ C {\bf 1}, 235 (1998).

\bibitem{MarquesdeCarvalho:1999ia} 
  R.~S.~Marques de Carvalho, F.~S.~Navarra, M.~Nielsen, E.~Ferreira 
  and H.~G.~Dosch,
  Phys.\ Rev.\ D {\bf 60}, 034009 (1999) 
  [hep-ph/9903326].

\bibitem{Liu:2009sn} 
  Y.~L.~Liu, M.~Q.~Huang and D.~W.~Wang,
  Phys.\ Rev.\ D {\bf 80}, 074011 (2009) 
  [arXiv:0910.1160 [hep-ph]]. 

\end{thebibliography}
\end{document}